# Application of Deep Convolutional Neural Networks for automated and rapid identification and characterization of thin cracks in SHCCs

Avik Kumar DAS[1], Chrisopher K Y LEUNG[1], and Kai Tai WAN[2]

[1] Dept. of Civil and Envionmental Engineering, HKUST, Hong Kong
[2] Xtrasensing Limited, Hong Kong

Abstract: Previous research has showcased that characterization of surface cracks is one of the key steps towards understanding the durability of strain hardening cementitious composites (SHCCs). Under laboratory conditions, surface crack statistics can be obtained from images of specimen surfaces through manual inspection or image processing techniques. Since these techniques require optimal lighting conditions, proper surface treatment, and prior (manual) selection of the correct region for proper inference, they are strenuous and time-consuming. Through this work, we explored and tailored deep convolutional networks (DCCNs) for the rapid characterization of cracks in SHCC from various kinds of photographs. The results from the controlled study suggest that the inference ability of the tailored DCCN (TDCNN) is quite good, resilient against epistemic uncertainty and tunable for completely independent but adverse observations. From the crack pattern computed using TDCCN, average crack width (ACW) and crack density (CD) can be calculated to facilitate durability design and conditional assessment in a practical environment.

1 Introduction

   Deterioration of concrete structures is a major problem with the civil infrastructure. The formation and opening of cracks are found to greatly facilitate the penetration of water and chemicals to accelerate the degradation process.  Strain hardening cementitious composites (SHCC) are a special type of fiber-reinforced cementitious composites that exhibit multiple cracking crack width  ~100 microns controlled intrinsically based on the micromechanical parameters of the mix (such as fiber, matrix properties) [1]. SHCCs have therefore been seriously contemplated for practical applications [2]-[4]. In Japan, SHCCs have been used for the construction of earthquake-resistant structural components such as dampers for tall buildings [5] and girder end slabs in bridges [5], [6];  as significant energy can be absorbed during the multiple cracking processes. After an earthquake (or a hazard) has occurred, a reconnaissance study needs to be performed to ensure the operational safety of the structure.  Maximum crack density ($CD_m$) of an SHCC is governed by the micromechanical parameter of the constituents. This will be discussed further in section 4. In this case, the statistics of surface cracks i.e. crack density (CD) can be used to assess the level of damage. A more damaged SHCC structure closer to failure is expected to have more cracks over the same region as compared to a less damaged one. The level of damage can, therefore, be initially screened from the cracking condition over a unit length. Apart from that SHCCs are also utilized as a  surface layer to provide a concrete cover with tight crack control [5], [7].  The latter approach is also applicable to existing members which need to be repaired or renovated to extend the service life. In order to develop a suitable mix for such application a large number potencial candiates need to be experimented with to gather the information of mechanical performance such as Young's modulus, and strength of the mix as well as durability performance of the mix. The durability performance is also governed by surface crack properties (such as average crack width) at different strain levels[8]. Intuitively, an SHCC mix with a larger number of crack with lower average widths will be more durable than another mix with a smaller number of cracks with larger average crack widths for a structure operating at the same strain level. Setting up of an vision based system to document crack formation at different strain level is possible however, visual inspection of each photograph to find the average crack width is extremely tedious. An experienced person (material scientist/engineer) can then exercise engineering judgement on whether a more detailed analysis is required (for a particular application) after results are automatically screened based on the average crack width. In summary, for both durability and resilience assessment of structures with SHCC members, there is a need to develop techniques that could automatically identify as well as to computation of crack parameters (such as calculation of crack location, density and average widths) for rapid screening and decision of follow-up action.

With the progress of structural health monitoring, many different methods based on various principles have been developed for damage monitoring. They can be classified based on the operating principle such as stress wave [9]-[14], electrical impedance/resistance [15], structural vibration [16], photographic techniques [17]-[19], strain gauges and fiber optics [20], [21]. Specifically for these problems, photographic techniques is more suitable because this could cover a large area with minimal permanently installed components and could produce a more direct result of the crack parameters (density and average width) [22]-[24].

1.1. Related Work

Image processing techniques have been used in the literature to detect cracks in reinforced samples [18]. The oldest technique includes (sobel/canny) edge detection techniques and adaptive thresholding techniques such as otsu's technique [18], [25]. Hoang et al. [26] showed that in order to improve the results of these conventional methods additional pre-processing needs to be done on the image such as enhancing the contrast. As compared to these conventional technique, other advanced technique were also introduced. O'Bryne et al.[27] and Gavilan et. al.[28] have demonstrated the use of machine learning (ML) model-support vector machine(SVM) to identify cracks whereas, Lee et. al.[29] and Liu et. al.[30] have demonstrated the applicability of feed forward neural networks (FNN) for doing the same. Recently, various researchers have used convolutional neural network (ConvNet) and found be more successful in detecting cracks where envronments are more complex [31]-[34]. In comparision to the cracks (of RC structures) of the above methodologies, the cracks in SHCCs has multiple thin (~100 micron). Due to this size of the cracks are comparable to the surface undulations/artifacts (see Figure 2) thus, adversely affecting the detection of SHCC cracks. As a result, SHCC cracks are more complex and

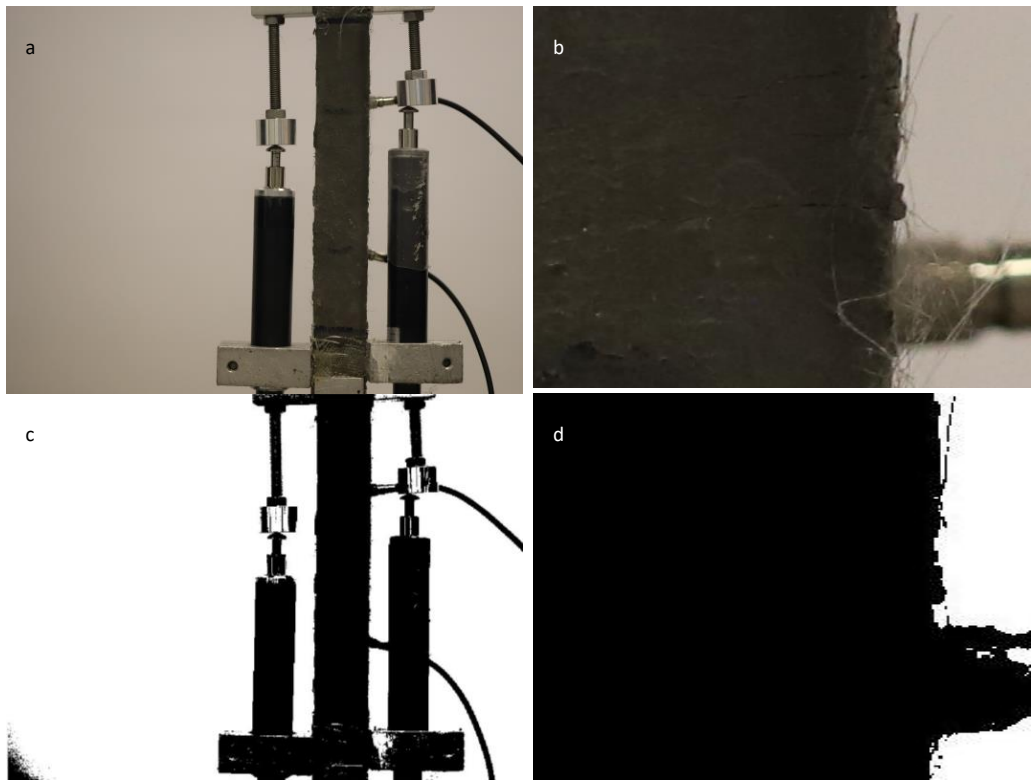

*Figure 1a) An example of typical test image b) Zoomed in portion showcasing the cracks c) Binary image of Figure 1a d) portion of binary image corresponding to Figure 1B*

harder to detect.

Methods which shows promising results for SHCC cracks includes crack scope, manual detection of cracks from the images [35]. Both of these requires significant manual labour. Recently, a Lu et al.[36] proposed image processing technique with 2 thresholds to detect SHCC cracks . However, in order for this method to work a) the sample surface requires to be painted to remove suface undulation , and b) preselection of the sections which have cracks by researcher. This aspect is explained with the help of figure 1. Figure 1a shows a typical image of a sample (without any processing on the sample) in this study and cracks observed within a section of this sample is shown in figure 1b. Figure 1c shows the result after binarizing of the image (figure 1a) (without prior selection of the appropriate region) and figure 1d shows the part of the binary image in Figure 1c corresponding to the region shown in figure 1b. Expectedly, crack was not visible within Figure 1d which clearly demonstrate the limitation of current image processing technique for crack detection. This shows that their application in the field condtion may not yield good results where images would be subjected to large in-class variation caused by poor adjustment to lighting condition, environmental noise and/or surface distortion [34], [37]. Even in the labaatory condition before these image processing techniques are applied, the section of the image with cracks have to be pre-selected by the researcher/inspector. Thus, this is cumbersome and extremely time consuming from a large amount of data.

1.2 Problem Statement

Development of a technique for rapid detection and computation of crack parameters (average width, density) of SHCC cracks is essential for assessment of damage and durability in both laboratory and in practical condition. However, at present there does not exist a method to do so for SHCC in which there are multiple thin cracks which are comparatively harder to detect. Our contributions are:

- As per the authors' knowledge, this is the first paper which attempts to detect thin and multiple SHCC cracks utilizing ConvNet for a practical situation, because images are taken without prior bias on the optimal experimental settings (such as polishing the surface and painting with white/yellow or to artificially enhance the contrast and/or ensuring proper camera angle with appropriate lighting condition to only focus on the preselected region).
- A novel image processing technique is also introduced to compute crack parameters (crack density, average crack width) from deep learning results.
- A novel trained ConvNet (TDCNN) was developed which can be potencially utilised material researchers and practioners to detect thin and multiple SHCC cracks in presence of artifacts such as sensors (LVDTs) sensor wires, markings, uneven sample edges, background cluttered samples and presence of additional objects found in and around laboratory (such as setups used in loading, residue from glues) which has similar features as cracks.

In this paper, the possibility of utilizing the ConvNet to identify and locate surface cracks of SHCCs under practical conditions was explored. At first, various viable solutions including image processing techniques, shallow feed-forward neural network (trained from scratch), and tailoring (pre-trained) deep convolutional neural networks were compared to select the best classifier. After that, effect of random epistemic uncertainty on the quality of the prediction for the best classifier was investigated. Then, the possibility of learning a new (un-correlated) dataset using this classier was examined. These (controlled) studies illustrate the potential of tailored deep learning-based methodology for computing crack parameters. To showcase applicability in a realistic situation, the development of crack patterns in SHCC was studied using this deep learning methodology.

2. Data Collection and Methodology

2.1 Data Collection and Preparation

Different camera phones as well as DSLR camera (canon DX 6 with standard lens) were utilized to capture images during testing of SHCC specimens in our laboratory. The samples were deliberately subjected to different lighting conditions during the test leading to natural variation in the brightness and sharpness of the images. The images show surface cracks on SHCCs (including reinforced SHCCs) generated under various tests in our laboratory.  In all cases, the surface features (such as non-uniformity), makings by the pen and sensors (for displacement measurement) were kept in the image as they were photographed. Different imaging instruments (cameras) capture images in different resolutions (5MP – 20MP) which could lead to a potential incompatibility. Therefore, it is important to select a resolution of an image, which circumvents such incompatibility while ensuring the accuracy of the detection result. To this end, each RGB Image (I) is cropped into small RGB segments (Is) of 227x227 pixels. Such segmentation achieves a good compromise between computational cost and accuracy of the detection results while ensuring uniformity across the images captured by various imaging instrument used for photography. Thereafter, each image segment (Is) is annotated as cracked, i.e. Positive (P), or non-cracked, i.e. Negative (N), segments. A detailed description of these image dataset is reported in Table 1. This dataset (collection of image segments (Is)) is called natural cracks (NC). Some positive examples are shown in Figure 2a and negative ones in figure 2b.

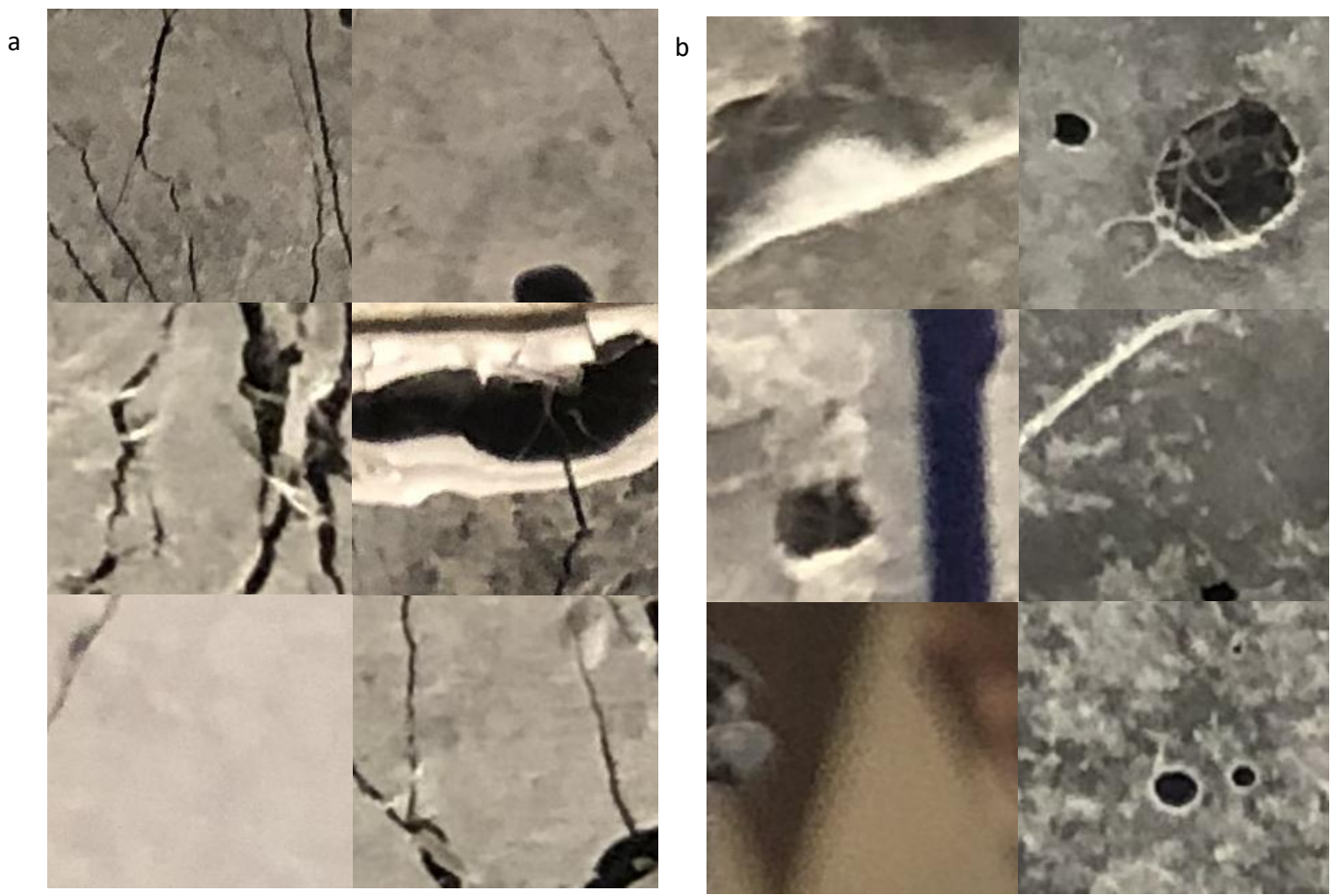

*Figure 2 some examples of a) Positive and b) Negative images of NC image dataset*

Table 1 Details of crack/non crack segments present in NC image dataset

| Image Dataset | Cracked Segment (P) | Non-Crack Segment (N) |
|---|---|---|
| Natural Cracks (NC) | 3112 | 2974 |

2.2 Methodologies

*2.2.1 Adaptive Thresholding (AdT)*

Image processing techniques such as thresholding /edge detection is used to segment cracks from greyscale images for SHCC [35]. For pertinent comparison adaptive thresholding technique (AdT) was also employed for NC dataset. In this method, the image segment (Is) was transformed to greyscale ($Is_g$) and then binary image was created employing Otsu's technique [25] to adaptively calculate threshold (Th) from the intensity distribution of $Is_g$. If the value of the ith pixel in $Is_g$ is below Th, it would be classified as crack (Eqn. 1). The existence of *crack pixels* in $Is_g$ leads to classification as positive (P).

$$crack = (Is_g^i < Th) \tag{1}$$

*2.2.2 SFNN*

In the second method, a shallow feedforward neural network (SFNN) was used to detect cracked images. Our proposed SFNN architecture consists of one input layer, two hidden layers and one output layer. The input layer consists of (m*n) number of neurons and each hidden layer had 128 neurons while the output layers has 2 neurons corresponding to 2 classes P and N. SoftMax function was then employed for probabilistic classification of image (segments) into a cracked or positive class (P) and a non-cracked or negative class (N). The architecture of this SFNN is shown in Figure 3a. The following cases were considered in this study:

a) Image segment (Is) was used as an input image for SFNN network. In this case, the input dimensions ($Is_{mxnx3}$) comprise m rows and n columns with 3 channels corresponding to red, green, and blue each. This case is referred as *SFNN-rgb*.

b) In the second case the (Is) was converted to greyscale $Is_{g(mxnx1)}$ and then used as input image for SFNN. There are still m rows and n columns in this case. However, the number of channels was shrunk to 1. This case is referred to as *SFNN-bnw*.

Apart from the input dimension every other aspect of the SFNN were kept the same.

*2.2.3 Tailored-DCNN (TDCNN)*

A neural network with more than two hidden layers is classified as deep network [38]. Deep networks are extremely powerful in autonomously learning the feature map of an input data set without the need of feature engineering [39]. Therefore, such methods have been employed to autonomously interpret large data set (to assist new discoveries) in many fields including medicine, astronomy and finance [40]-[42] . For datasets such as images and videos, deep convolutional networks have been extremely efficient. Their performance is comparable to a human in detecting (identifying) spatial features (and their inherent relationships) present in images such as patches/edges (of similar shape, texture or color) leading to superior image classification capability [43]-[46]. A typical deep convolutional network consists of the 3 types of computational units: a) convolution unit which generates feature maps through spatial filter b) activation unit to (non-linearly) select complex feature maps and c) pooling unit to down-sample selected image and features, that enables dealing with multiple resolution of the spatial feature present in the image [38]. It is possible for researchers to 'engineer' DCNN from scratch by iteratively adding these computational units [47]-[49]. In the case of SHCC which is a new material, datasets of images of the cracks are limited. This may lead to overfitting, so the resulting classifier may not be resilient. Therefore, such idea is not feasible for separation of SHCC cracks. This problem can be circumvented by using transfer learning [38], [50]. The key of transfer learning is to repurpose a learnt representation of common features within images in large and general dataset and then utilizing it for a specific dataset such as classification of SHCC images [38], [50]. Such transfer work when the base (from where the features are learnt from) is extremely large so that a general representation of the features of images could be learnt [50]. Therefore, choosing a DCNN that showcase state of the art performance in large dataset such as 'ImageNet' is essential to create a resilient crack-image classifier. Specifically, ResNet, an established DCNN which has been successfully used in other civil engineering application [51], is chosen as the base for further fine tuning.

Out of all the variations of ResNet, the architecture of ResNet18 was selected and trained with images in 'ImageNet' database. After the training was complete, and the architecture (including the weights) of this network were concatenated with output layer of SFNN as shown in figure 3b. This tailored network is hereby referred to as TDCNN, and its architecture is shown in Figure 3b.

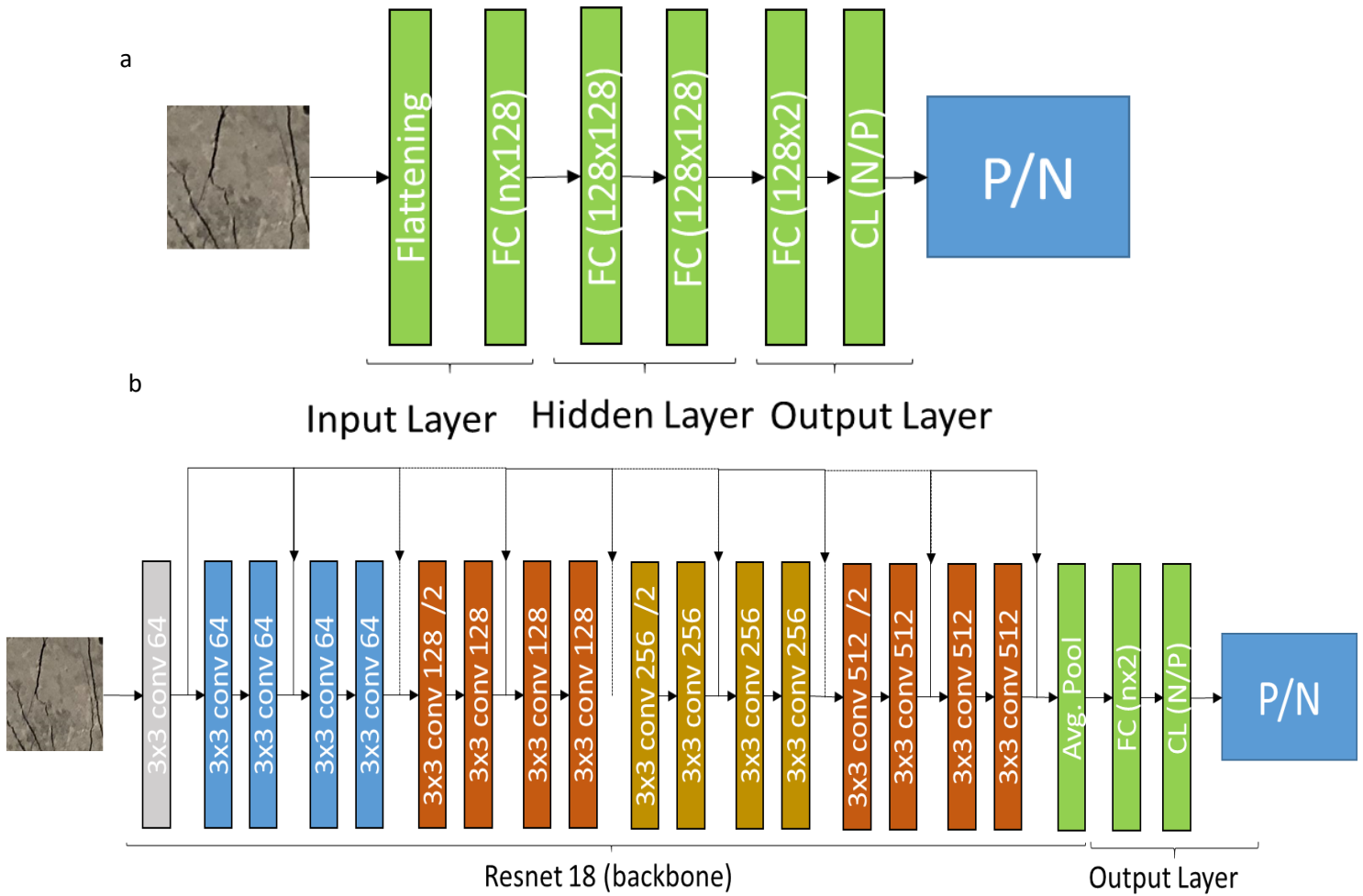

Figure 3 Architecture of a) SFNN-rgb  b) DCNN

3 Results and Discussion

3.1 Training Process for neural networks (NNs)

Unlike AdT, techniques based on neural networks i.e. SFNN-rgb, SFNN-bnw and DCNN require training on NC dataset. To prevent overfitting during training, the dataset is divided into 3 sets. These are training, validation, and test sets. Neural network learns the characteristics of the data during training over training data set. The optimal parameters of these neural networks are computed by iteratively minimizing the (cross entropy) loss (L) (Eq. 2).

$$L = -\sum_{i=1}^{M} y(o,c)\log(p(o,c)) \qquad (2)$$

Here, M is number of classes, $y(o,c)$ is binary value either 0 if predicted class label (c) does not match with observed data(o) or 1 otherwise and p (o,c) is the probability of the predicted class label (c)

Unfortunately, training of neural networks is not straight-forward. Several factors (hyperparameters')

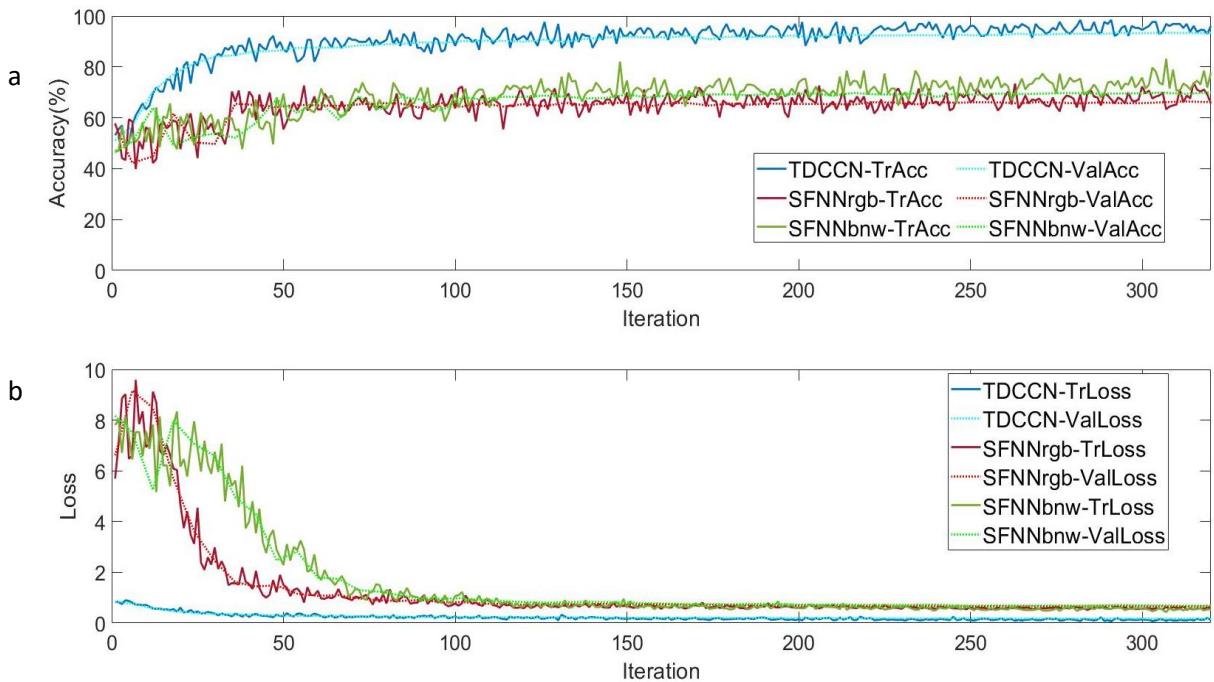

Figure 4 Evolution of a) accuracy and b) loss for NNs in first few epochs

such as epochs and choice of optimization algorithm affect the final accuracy of the trained network. The validation set was used to decide the optimal choice of these hyper-parameters. Model evaluation was performed on the test set. With 10% of the total data reserved as test data, 90% was used as training+validation (tr+val) sets, out of which 0.7 times were randomly selected as training set and the rest as validation set. The evolution of loss and accuracy for the first 2 epochs (each epoch corresponding to ~160 iterations) for NNs is shown in Figure 4.  In all NNs, as the number of iteration increases, loss function goes down, and (prediction) accuracy in the validation set increases but plateaus within 50 iterations. This shows NNs are successful in learning the structure within the dataset.  When compared to SFNNs, DCNN reaches the plateau earlier and has lower loss value. This is due to the pre-training of this network on "ImageNet."

3.2 Comparative Analysis of different techniques

In this section, the performance of methodologies such as conventional techniques -AdT, Contrasting Image+ Otsu's Technique [26], Machine Learning Technique- Support Vector Machines(SVM) [52], [53] ,and Random Forest (RF) [54], Neural Networks-SFNN-rgb, SFNN-bnw, ,and TDCNN were systematically studied. Their performance was evaluated to select the most promising technique for further investigation. To this end, NN models were trained to classify image dataset $(Is_{1...n}^{NC})$ with n images of natural cracks (NC). For each image ($Is_i^{NC}$) in NC dataset, predicted probability ($Y_{pred}$) of different classes (N or P) is evaluated. Based on the highest probability, each image is classified into the most probable class. Indicators' accuracy (acc), precision (pre), F1-score (F1)  were calculated to evaluate the model. The calculation procedure for these parameters is detailed in Appendix 1. The mean values of these parameters calculated for NC datasets along with network complexity are reported in Table 2. Here, the number of parameters of the model was used as a representative for complexity.

*Table 2 Performance evaluation of different methodologies*

| Technique | Methodology | No. of Parameters/ Complexiety | acc (%) | pre | F1 |
|---|---|---|---|---|---|
| Conventional Techniques | AdT | ND | ~50 | *** | *** |
|  | AdT after image contrast [26] | ND | ~52 | *** | *** |
| Machine Learning | SVM | O(n^2) | 58.9 | 53.2 | 54.1 |
|  | Random Forest | O(nlog(n)kd) | 68.9 | 62.1 | 64.3 |
| Neural Networks | SFNN-bnw | ~6.6 million | 71.4 | 64.6 | 64.8 |
|  | SFNN-rgb | ~19 million | 68.2 | 60.4 | 63.3 |
|  | TDCNN | ~11 million | 91.4 | 88.3 | 89.4 |

ND is Not Defined    *** low therefore Not Evaluated   k=number of tress (~150) d=dimension of image (227x227x3), n is the number of training examples

Apart from the complexities introduced by uneven lighting captured at different angles (changes in the viewpoint), our dataset is filled with artifacts such as sensors (LVDTs) sensor wires, markings, uneven sample edges and presence of additional objects found in and around our laboratory (such as setups used in loading, residue from glues) which has similar features as cracks. These aggravates already large inter-class variation. Expectedly, a combination of these factors leads to inferior performance of conventional methodologies such as AdT. As many of these factors (lighting condition, physical environment) are not the same or known apriori, using AdT is expected to be cumbersome and time-consuming. However, in cases where cracking area is pre-selected, lighting conditions are uniform, and stability of the camera hardware is assured, AdT could still be used. As compared to conventional techniques, both machine leanring (ML) and feed forward neural networks (FNN) perform superior in detecting SHCC cracks. This is because the NN can learn the variability within the image whereas pre-coded feature (in conventional techniques) is extremely 'brittle' and breaks down when there is a large variation within images. It is interesting to note that SFNN, there are 3 channels (rgb) so the number of parameters is substantially higher as compared to the base with bnw. As a result, computational complexity for training *SFNN-rgb* is higher than *SFNN-bnw*. Even with lower computational complexity, the performance SFNN-bnw is slightly better than SFNN-rgb; this might corroborate that cracks are relatively easier to detect from a greyscale image compared to their RGB counterpart. For TDCNN, accuracy is highest as compared to other methodologies even though the number of parameters used for training is fewer than FNN and ML based technique. This is because FNNs and ML based techniques perceive each pixel independently which (completely) breaks down the (2D) structure (such as patches) within an image. However, unlike FNN and ML based technique, TDCNN preserve and learn from the spatial structure, so it is possible to learn (discover) high level features that distinguishes an image with crack from a non-crack image even when there is large in-sample variation.

The results from comparative analysis shows that due to large inter-class variations in the SHCC datasets, other techniques (such as conventional, ML and FNN) apart from the proposed TDCNN is not effective in discriminating SHCC cracks from the background. The study also shows that TDCNN could be the most attractive solution for autonomously segregating crack segments (P) from non-crack segments (N). Therefore, in the next sections, properties of TDCNN will be investigated further.

3.3 Resiliency against epistemic uncertainty in image quality

Civil infrastructures based on SHCCs are designed to last for at least 50 years. It is possible that imaging sensor installed on the structure may change over its life span and sometimes it may also fail to capture sharp images of the cracks. Therefore, it is important to verify that TDCNN could also learn such conditions as variations within the image and could still segment the cracks. In this section, effect of such adverse epistemic uncertainty in image quality on the classification accuracy of TDCCN was comprehensively studied. Reconnaissance study in a post hazard situation usually employs Unmanned Aerial Vehicles (UAVs) for collection of images. In those cases, quality of the images might be affected environmental factors. These environmental uncertainty affects could be caused by the stability (mal-functioning) of the imaging instruments in UAVs leading to subjects may not be in focus i.e. images are blurred or images are interlaced with noise i.e. Noised. In other cases, depending on the site-condition surface damages (cracks) might be partially blocked i.e. Hidden or in dusty condition, lens might be partially covered with dust the leading to lack of color/contrast in the images. To study these effects, NC dataset was digitally modified. An example highlighting digital transformation of a typical image is shown in Figure 5. The parameters used in the digital modification are reported in Table 3. 1200 positive (P) images from NC dataset which have cracks near the center were digitally modified. Because such modifications were random, image with cracks only at the corner might disappear due to digital modifications leading to inaccurate "ground truth". Similarly, 1200 negative (N) images were also digitally modified to ensure the total number of P and N for the training remains the same.

*Table 3 Parameters used for digital modification*

| Type of Modification | Noise | Hidden | Blurred | Lack of Color/saturation |
|---|---|---|---|---|
| Parameters | 'salt and pepper' random, 0.15 | X, Y direction each rand between (0, 0.2) each | Gaussian Blur random between (0,3) | Color changed to greyscale and saturation changed |

|  |  |  |  | randomly between [-0.5 1.5] |
|---|---|---|---|---|

The new natural crack dataset (NNC) consists of both NC and digitally modified NC images. Following the process described in section 3.1, TDCNN was trained for NNC dataset. The performance parameters calculated for this case are reported in table 4. Figure 6 shows the receiver operating characteristics (ROC) curve for both P and N classes. (Note: Derivation of ROC curve is discussed in appendix 2.) More specifically, the area under the ROC curve (AUC) is a measure of the separability of a classifier [55]. Higher the area, the better the model performs in distinguishing images accurately into N and P classes. An exceedingly high (>0.9) AUC was observed for this case, indicating that TDCNN is incredibly good in segregating images into the different class (even when in-sample variations are remarkably high).

*Table 4 Performance of DCCN on several datasets*

| Dataset | Accuracy (%) | Precision (%) | F1-Score |
|---|---|---|---|
| All combined (NNC) | 93.1 | 91.2 | 91.8 |
| NC | 91.6 | 90.4 | 90.2 |
| NC-Noised | 88.6 | 86.4 | 86.7 |
| NC-Hidden | 96 | 93.9 | 85.6 |
| NC-Lack of color, contrast | 98.5 | 97.2 | 96.8 |
| NC-Blurred | 94.9 | 92.2 | 93.4 |

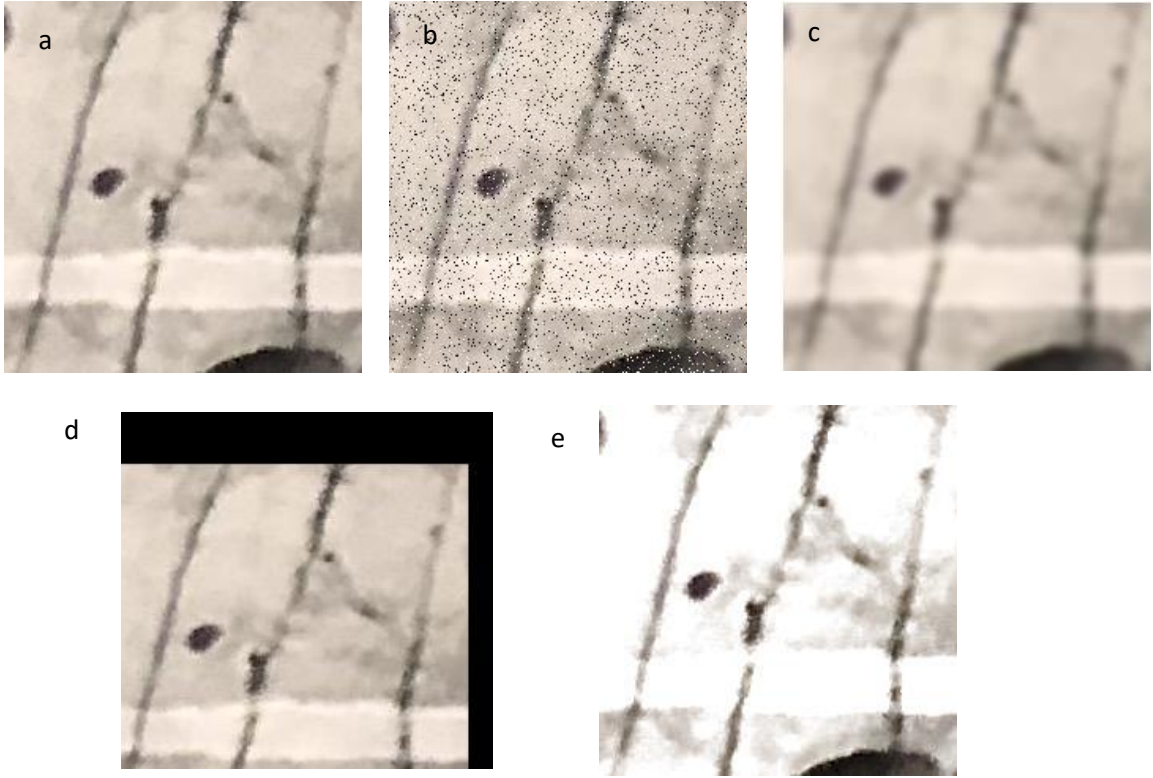

*Figure 5 Example showcasing digital modification a) original image b) Noised Image c) Blurred d)Partially Hidden and rotated e) Modified color and contrast of original image*

The result in Table 4 shows that the accuracy of the trained TDCCN slightly increased for NNC (the digitally modified dataset) as compared to NC dataset, as there are more data for TDCCN to learn the underlying structure leading to better segmentation. This result is very important as it shows that with every cycle of monitoring, even if the data collected in subsequent cycles have different types of epistemic uncertainty, this uncertainty will not adversely affect accuracy (performance) of TDCCN.

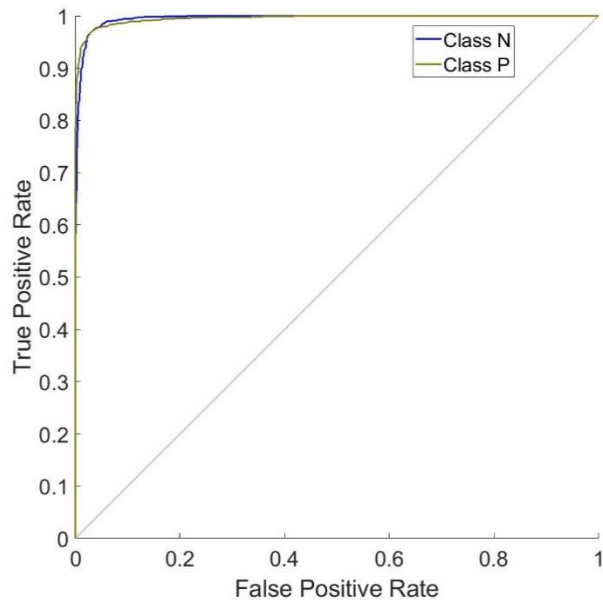

*Figure 6 ROC curves of classes N and P calculated for NCC dataset*

Not every type of uncertainty has the same effect, so it is important to understand which kind of uncertainty has the most adverse effect. This can help researchers and practitioners plan data collection process for SHCC cracks in the future. Ablation study was performed for different classes of images in NNC dataset a) NC (test data consists of only NC) b) NC-X were performed. Here, 'NC-X' is the test data consists of only X type, where X is the type of modification made on the original image (e.g. blurring). The (mean of) performance parameters calculated from the trained network for these subsets of the images is also reported in Table 4. It was observed in all cases the performance of TDCNN was acceptable (>85% accuracy), with the worst performance observed for NC-noised images. This could be because 'salt and pepper' noise (as shown in Fig.5b) leads to sharp and sudden change in the pixel value of the images leading to discontinuity.

3.4 Adaptability to a new (adverse) image dataset

TDCNN can autonomously learn the features to distinguish images containing cracks from non-cracks with remarkably high accuracy for NC and NCC data set. It should be noted that NNC dataset is created by digitally modification of NC dataset. To showcase more comprehensive applicability, it is essential to demonstrate that TDCCN could still (learn to) segregate crack images for a new dataset that is uncorrelated to NC. Typically, humans can identify the presence of an element (such as cracks) within an image by the contrast between such patch and its background. One such important clue is the luminosity (or lack of it in the crack). By cluttering the background with similar clues would make cracks harder to identify and therefore significantly affects the performance of a human operator. Cluttering can be created on specimen to mimic a digital noise. With respect to the results reported in Table 4 , such cluttering is expected to represent one of the worst conditions for TDCCN. In this section, the possibility of learning a new dataset of SHCC specimens that is adverse and un-correlated to NC was examined. Images of SHCC specimens with surface cluttered (speckled) with black paint were collected. The speckles make the cracks relatively hard to search manually. This dataset is called speckled cracks (SC). In this dataset **3852** image segments were annotated as positive while **7876** segments were annotated as negative manually. Figure 7 shows some examples of positive and negative images. Similar to the process described in section 3.3, 1200 images each are randomly selected from positive and negative within the dataset were digitally modified to create larger in-class variability. The parameters used for digital modification is reported in table 3. The new dataset containing original and digitally modified images is hereby referred to as the SCC dataset. For training and benchmarking, N annotated images were randomly selected from SCC/SC database to match the number of P annotated images within that dataset. The 'box and whisker' plot in figure 8 shows the results for SC and SCC datasets. . The mean (black dot) and maximum (cross) for each performance parameter is also shown in Figure 8. The results indicate that the mean accuracy of SCC is (slightly) higher than the SC dataset. This might be because there are more images in the SCC dataset to facilitate learning. The inference ability of TDCNN in this worst-case scenario remains reasonably accurate with mean accuracy of ~90% and maximum accuracy of approximately 94% that is remarkably close to the human level performance. As compared to an SCC dataset, the SC dataset has slightly lower variability in performance (as changes in the random selection do not significantly alter the in-sample variability of the samples as compared to SCC samples. For both datasets, the accuracy for all cases is >85%. These indicative results demonstrate that TDCCN is applicable for a completely different and adverse situation.

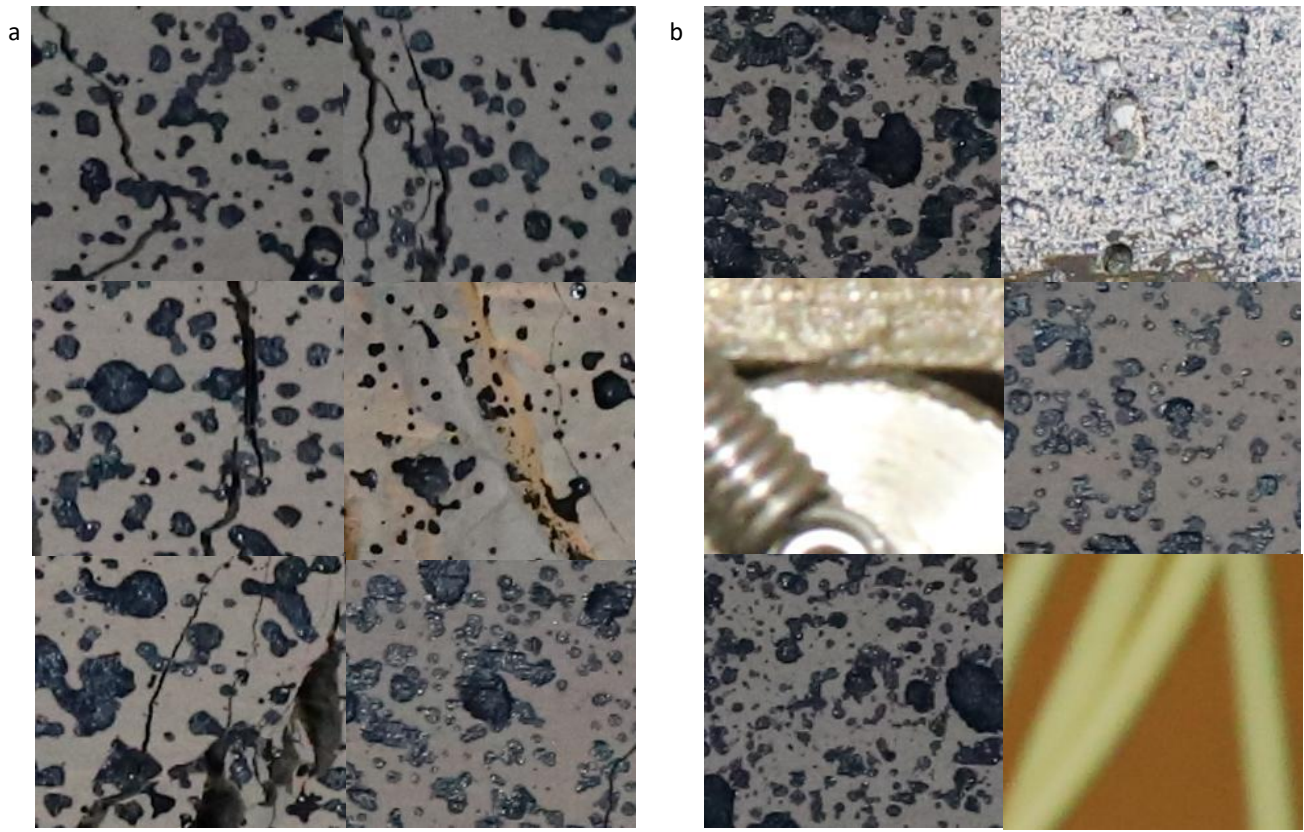

*Figure 7 Examples of a) Positive b) Negative image segment from dataset SC*

4. Applicability of TDCNN for studying crack pattern development in SHCC

Through controlled study, TDCCN is found to be superior as compared to other techniques. It is still necessary to showcase that TDCCN can supply useful information in practice. TDCCN was therefore employed to trace crack propagation in SHCC coupon specimens under tensile stress. This test achieves a good compromise between computational cost/accuracy and applicability of the proposed technique. Moreover, the behavior of cracks (crack pattern, crack mechanisms) under applied tensile loading has been well established in the literature [14], [56]-[59]. Consequently, it is relatively easier to benchmark the results from TDCNN with those in the literature.

## 4.1 Mix Design and Testing Procedure

The objective here is to generate cracks to demonstrate the potential of TDCCN in a non-controlled setting. The SHCC mix used for this study is reported in Table 5 and satisfies the multiple cracking criteria [60]. Following the standard procedure for mixing the materials [13], [14], SHCC coupons (S1,S2,S3) of 350mmx25mmx8mm were casted. These specimens were placed in a mold, kept wet for 2 days before demolding and then placed in a curing chamber with relative humidity of ~95% for 26 days. After the curing was complete, aluminum plates of length 100mm were attached at each end to ensure proper gripping. These specimens were subjected to increasing tensile strain with 0.2mm/min loading rate. To ensure the results are representative of the real-world case, sensors and markings on the sample were kept as they are during the test and the use of external lighting during photo taking was avoided. In autonomous monitoring, the camera is relied upon to focus on the subject. To replicate this condition, imaging instruments were set to autofocus. As a result, the sample looks dark and is not always sharp (as shown in figure 8, 10).

*Table 5 Matrix Design for SHCC*

| Sample | Class F Fly Ash | OPC-53 cement | Silica Fume | Sand | Water | Super Plasticizer |
|---|---|---|---|---|---|---|
| S1,S2,S3 | 0.8 | 0.18 | 0.02 | 0.2 | 0.22 | 0.003 |

## 4.2 Computation of Crack Statistics

In computer vision, different tasks such as *identification* and *localization* have different levels of operating complexity. Searching for the correct frames in which cracks are present is an example of *identification* task. In this test, *identification* selects the frame in which cracks are present. An example of a (half) frame with cracks (positive *identification*) from our experiment is shown in figure 9a. This frame originally encompasses a physical space of ~0.2m$^2$ area. Unfortunately, this result is not enough for computation of crack statistics. Whereas, *localization* is the process of identifying the position of the crack within the frame. A segment of the frame shown in figure 9b where the crack is *localized* within each white box has an area of ~1cm$^2$. Clearly, a technique that localizes the crack has already identified the crack. Finding the location of (all) the cracks within a picture enables an estimation of approximate crack density, number of cracks and average crack width. As discussed previously, these parameters could be (qualitatively) used for characterization of SHCC mixes or the rapid screening of regions that are severely damaged in a hazard. Therefore, in this section, the focus is on showcasing potential applicability of the proposed technique to find the location of the cracks.

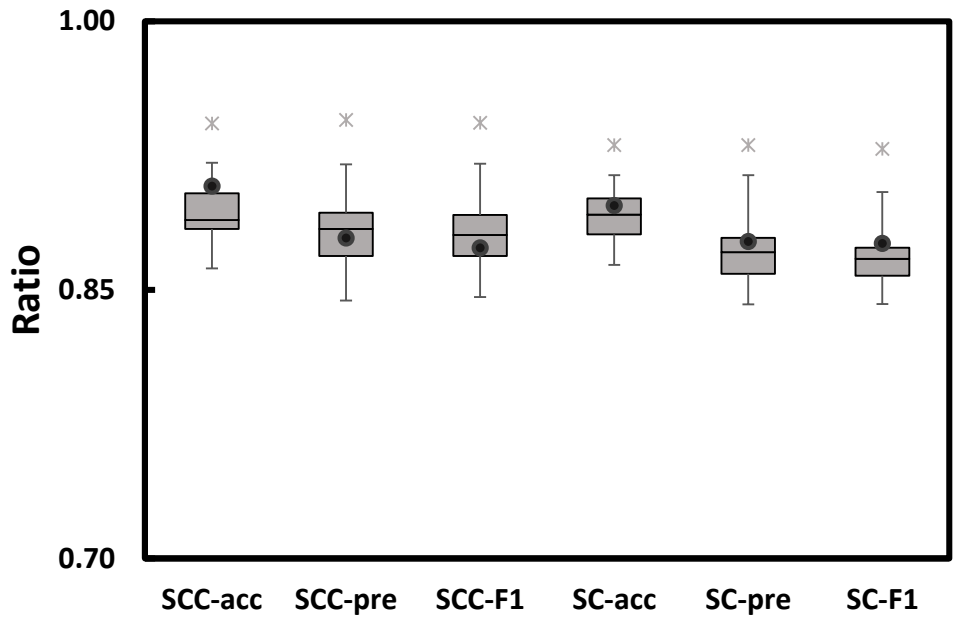

*Figure 8 Performance of DCCN on SCC and SC dataset*

A window search technique was employed to identify cracks within the image. That is, a window was slid across the image to identify the positive segments (P) and negative segments (N). The sliding length for the window is the same as that of the window size. This is schematically

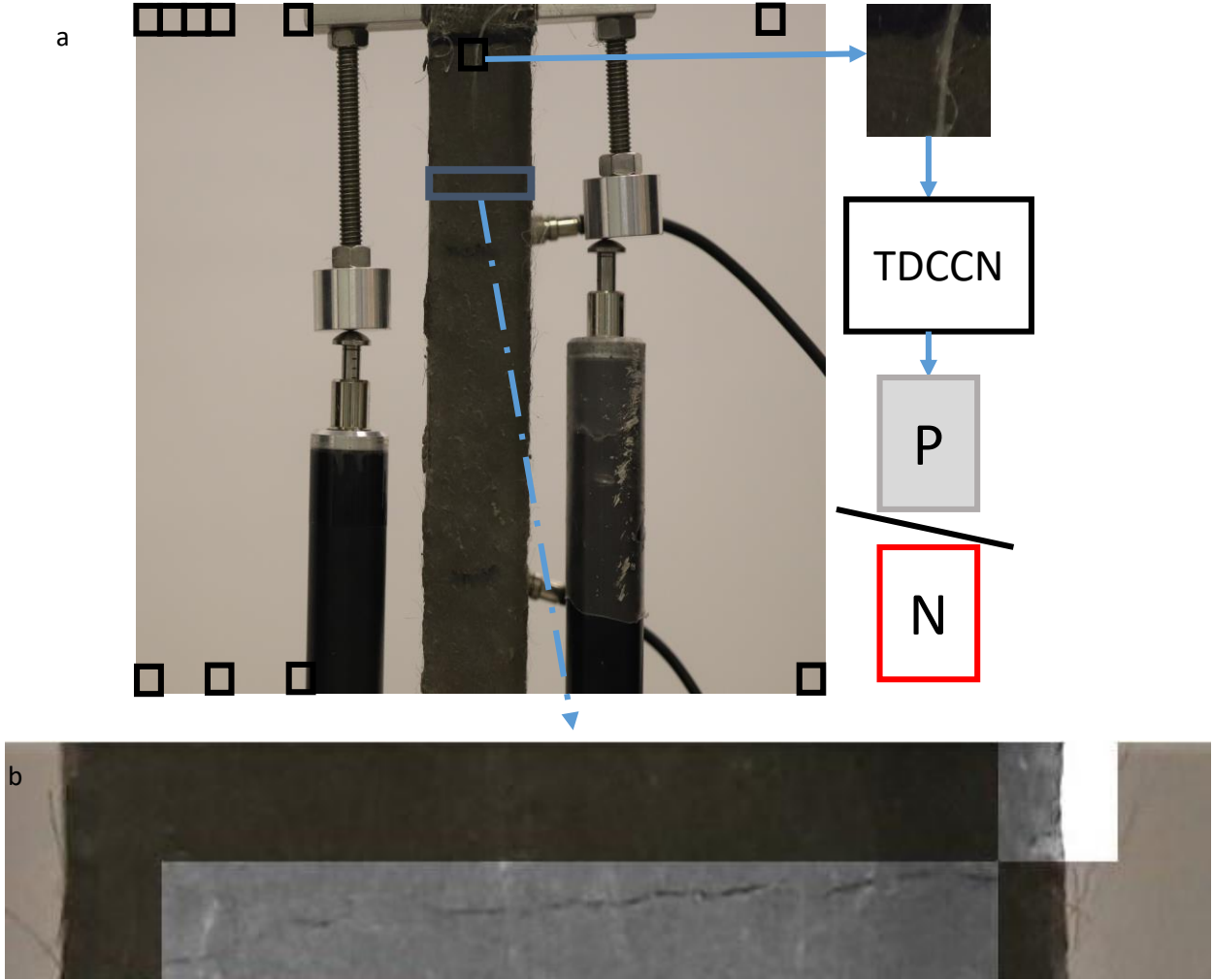

*Figure 9 a) Schematic of assigning positive or negative b) an example of cracked region computed through DCCN Note: crack region is enhanced and overlaid within the original image for better visualization*

shown in Figure 9a. By grouping the neighboring windows which are P, the location of the crack was found as shown in Figure 9b. The total area covered by adjacent white boxes (in figure 9b) is referred to

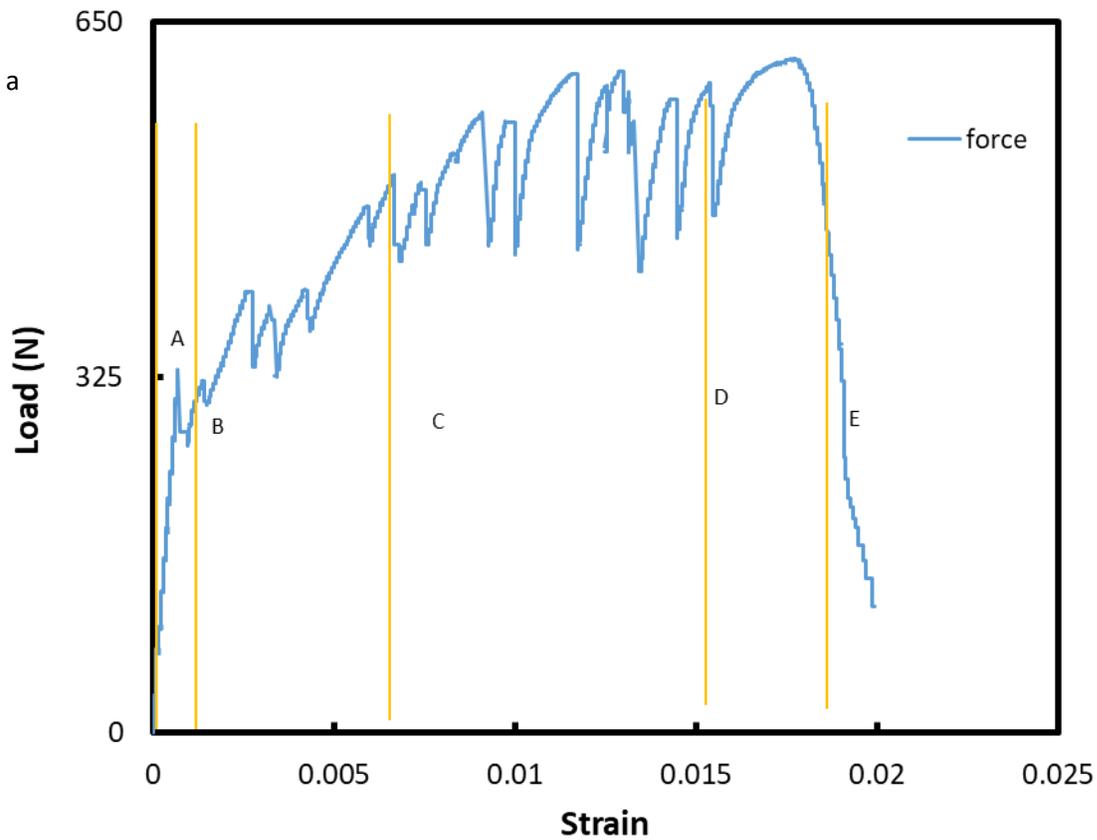

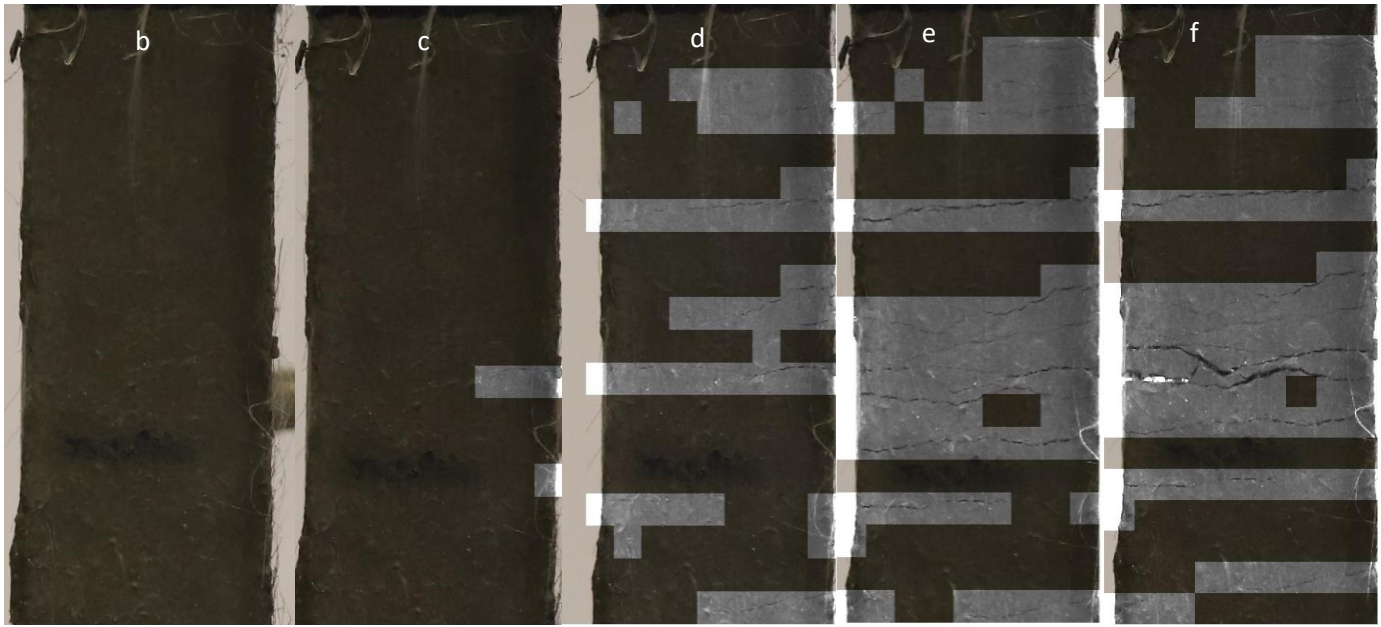

*Figure 10 a) Load vs Strain response b-f) Representative images showcasing development of cracks measured using DCCN*

as *localized cracking zone* (LCZ). Since each LCZ element is small (in terms of area) conventional image processing techniques could be used to enhance the image. To better demonstrate this, enhanced image was overlaid on the original image for LCZ as shown in Figure 9b. From this enhanced image, crack statistics such as crack pattern, crack number, average crack width (ACW) and crack density (CD) were estimated. This process was repeated for all the images collected during the SHCC tension test. The details of the calculation of crack number and the crack pattern are discussed in appendix 3.

4.3 Results and Discussion

*4.3.1 Crack Pattern Development*

The load vs strain curve of the specimen is shown in Figure 10a. The load corresponding to the frame number extracted for figure 10b is also marked in Figure 10a. Cracks in SHCCs are very small, and by looking at the whole frame (as apparent from figure 8) it is not possible to visually confirm the results of TDCNN. To this end, instead of showcasing the whole frame in figure 10b, the focus is placed on the approximately top 1/3$^{rd}$ of the SHCC sample. The location of the cracks in each frame was estimated following the method illustrated in figure 9a using TDCNN. Figure 10b-f shows the indicative results of the development of cracks on SHCC using 5 frames. In all these frames, DCCN can correctly localize the cracks within the frame with an average accuracy of 0.91. In these frames, the LCZ is enhanced and overlaid on the original picture for better visualization. The first macroscopic crack does not appear until tensile loading crosses the first cracking strength of the SHCC composite. This initial zone is referred to as elastic zone. Figure 10b captured at the beginning of the test expectedly has no cracks. Just after the turning point of the load vs strain curve, the first crack is visible as observed in figure 10c. This is the beginning of the hardening zone. In the hardening region, fibers bridging the crack transfer the load back into the matrix through the interfacial bond to induce additional cracking. In a properly designed SHCC, this process will repeat multiple times leading to the formation of multiple subparallel cracks on the matrix. This is observed through the crack pattern shown in Figures 10d and 10e. In comparison, the crack pattern remains the same in Figure 10f as compared to figure 10e but wider cracks were observed. This is because the strain corresponding to figure 10e is very close to crack localization after which strain softening would occur. As indicated in Figure 10a, after the last crack is formed (at point D), the load drops and then increases beyond the previous maximum (so the average crack width increases) until crack localization occurs with one of the cracks showing softening behavior. The observation of the crack pattern matches well the theoretical expectation and other experimental observations of previous researchers [14], [58], [59].

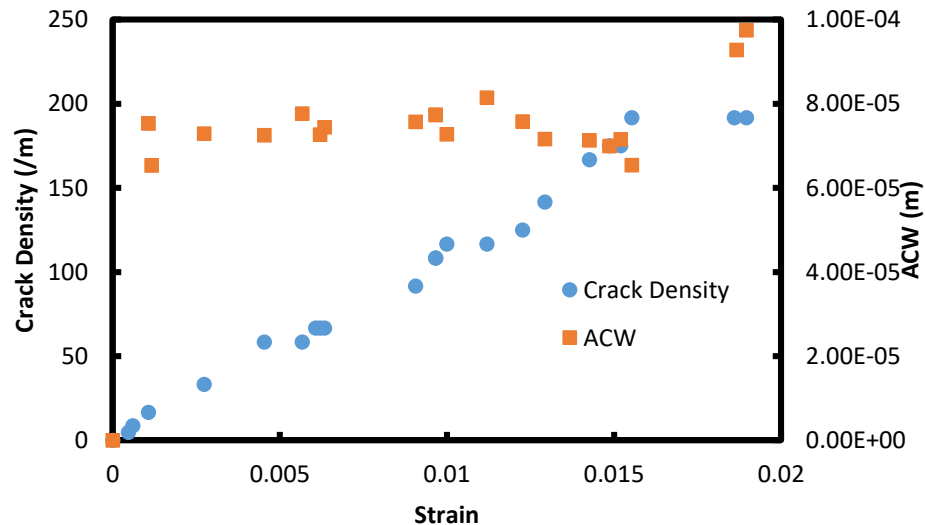

*Figure 11 Development of average crack width and crack density with applied strain for sample S1*

*4.3.2 Comparison of theoretical results of ACW with those estimated by TDCNN*

In an ideal SHCC with no variations in matrix strength within the specimen, when stress reaches to first cracking strength of this material multiple crack forms simultaneously. In this condition, all the cracks will have exactly the same crack width. The total number of such cracks will be controlled by transfer distance calculated through multiple cracking theory of brittle matrix composites [56]. As all the cracks have same crack width that is, theoretical average crack width ($ACW_{th}$) will remain constant between first cracking strength of the material until failure of the composite. This suggests ($ACW_{th}$) has value 0 when stress is less that first cracking strength and constant after first cracking strength and failure of the composite. However, in a realistic scenario the subsequent cracks in a SHCC forms at a higher stress than the previous crack. As a result, ACW is expected to increase as stress increases within the hardening region. Assuming, the slope in the hardening region is small linear constant model for ACW therefore employed.

The evolution of ACW in for sample S1 is shown Figure 11. Here, ACW is calculated from the total number of cracks (determined as in Appendix 3) by Eq.3.

$$ACW = \frac{LVDT\ displacement}{total\ number\ of\ cracks} \tag{3}$$

It was observed that there is a large deviation in ACW especially at the beginning of the test, which could be due to the time at which the photograph was captured. This can be explained with respect to Eqn. 3. In the beginning when the crack number is small, depending on whether the photograph was captured before or after a crack was observed, the denominator of Eq. 3 significantly changed which is observed in ACW. During the hardening region, ACW for this sample remains around approx. 75 microns. After the sample has failed the ACW increases rapidly and reaches close to 100 microns just after crack localization. This is expected, as localization leads to the rapid widening of one of the cracks, which significantly increases the average crack width. Specifically, for this sample fitting the experimental observation with linear constant model reveals that ACW as 77 microns with coefficient of determination of 0.92. This result of this study is in good agreement with theoretical expectation. Similarly, fitting for other samples (S2 and S3) in this study with linear constant function reveals average ACW for these samples as 72 micron.

Using this methodology based on TDCNN, ACW of a SHCC mix can be conveniently obtained in laboratory conditions from experimental images. As discussed previously, ACW is qualitatively correlated with the maximum crack width of an SHCC specimen which controls the transport property in turn durability of the specimen. To select a more durable SHCC mix, SHCC mixes in laboratory conditions can be screened based on the ACW. This method can potentially save precious time and labor of the researchers which is otherwise consumed screening and benchmarking through large numbers of SHCC mixes.

*4.3.3 Comparison of theoretical results of CD with those estimated by TDCNN*

As discussed in the previous section, ACW could be obtained from laboratory specimens to screen SHCC mixes for durability applications. Unfortunately, in a practical situation such as reconnaissance study after a natural hazard has occurred, it is often not possible to measure the relative displacement or strain over a certain part of the structure as there may not be pre-installed sensors/instruments and human assess (especially after a hazard) may be restricted. Consequently, it is not possible to calculate the average crack width. However, with the use of Unmanned Aerial Vehicles (UAVs), it is possible to take photographs of various parts of the structure. In such cases, crack density (CD) in a certain region can be calculated by dividing the total number of cracks by the length of the region, as in Eq.4.

$$CD = \frac{Crack\ Number\ (N)}{Length\ of\ the\ region\ (L)} \tag{4}$$

In this section, the trends of CD predicted by TDCNN is compared against results from the theoretical modeling of SHCCs under tensile loading. SHCC is micro-mechanically modeled as an ideal brittle matrix composite with three-dimensional randomly distributed discontinuous fibers. The theoretical lower limit and upper limit of crack spacing in such composites are x' and 2x' respectively [56], with x' calculated according to Eq.5 [56], [57].

$$x' = \frac{L_f - \sqrt{L_f^2 - 2\pi L_f \lambda x}}{2} \tag{5}$$

where $x = \frac{V_m E_m \epsilon_{mu} r_f}{V_f 2\tau_{eff}}$ \hfill (5.1)

$$\lambda = \frac{4}{\pi g} \tag{5.2}$$

$$g = 2 * \frac{e^{\frac{\pi f}{2}} + 1}{4 + f^2} \tag{5.3}$$

In the above equations, g is the snubbing factor, f is the snubbing coefficient, $L_f$ is the length of the fiber, $V_m$, $V_f$ are volume fraction of the matrix and fiber, $E_m$ is the Young's modulus of the matrix, $\epsilon_{mu}$ is the failure strain of the matrix, $r_f$ is the radius of the fiber and $\tau_{eff}$ is the effective bond strength of fiber-matrix interface

Based on x' obtained from Eq.5, the lower theoretical limit of crack density ($CD_m^{th}$) over a given region of fixed length L is estimated by Eq. 6. Eq. 6 suggests that the (max) CD of an SHCC is only dependent on micromechanical parameters of the constituent SHCC. Trend predicted by this result suggests that if strain of SHCC is less than the first cracking strength ($\epsilon_{cr}$) then CD is going to be constant (=0). As discussed above, the crack density would reach a maximum value before the ultimate tensile strength of the SHCC is reached. In other words, there exists a limiting strain of crack formation ($\epsilon_{lcr}$) beyond which CD is going to stay constant (at a plateau value $CD_m^{th}$). For strain ($\epsilon$) values between ($\epsilon_{cr}$) and ($\epsilon_{lcr}$), the SHCC is under tensile strain hardening, and the strain is due to the contributions from crack opening ($c_i$) and elastic strain ($\epsilon_{elastic}$) as in Eq. 7. As the elastic strain is much smaller than the strain caused by crack openings, it can be neglected and the strain can then be calculated by Eq. 8.

$$CD_m = (\frac{L}{2x'})/L \tag{6}$$

$$\epsilon = \sum_{i=1}^{N} c_i / L + \epsilon_{elastic} \tag{7}$$

$$\epsilon = N/L * avg(c) \tag{8}$$

Here, N is the number of cracks and N/L is the crack density.

In an ideal case where there is no variation in properties among various sections of the SHCC, all cracks will form simultaneously (at the same applied stress) and the width of each crack is constant. Under this situation, the strain and crack density are linearly related to one another. Under this ideal situation, the crack density (CD) is a trilinear function of the applied strain, staying at zero for strain less than $\epsilon_{cr}$, varying linearly between $\epsilon_{cr}$ and $\epsilon_{lcr}$, and staying constant for strain greater than $\epsilon_{lcr}$. In reality, due to the variations of inherent flaw size and fiber distributions within the SHCC, the number of cracks can only increase with increasing strain and the width of each crack also increases as strain is increases (because stress is increased correspondingly). The CD versus strain relationship within the hardening regime is therefore not exactly linear. Nevertheless, the linear trend still appears to be a good approximation of SHCC cracking behavior observed in experiments, as shown in [61]. If the tri-linear function is applied to the test results in this study, good agreement can also be achieved. Evolution of CD for sample S1 is shown in Figure 11. Specifically, for the experimental results of sample S1 shown in Figure 11, tri-linear fitting has a very high (>0.95) coefficient of determination ($R^2$). The computed $\epsilon_{cr}$, $\epsilon_{lcr}$, and $CD_m^{ex}$ were 0.00012, 0.0159 and 194 /m respectively. Figure 12 presents the tri-linear model estimated from experimental observation of samples S1, S2, and S3. In figure 12 the strain is normalized to ensure the maximum experimental CD ($CD_m^{ex}$) of each sample is reached when the normalized strain is '1'. The results indicate that average and standard deviation of $CD_m^{ex}$ for these samples were 186.33 and 16.03 respectively.

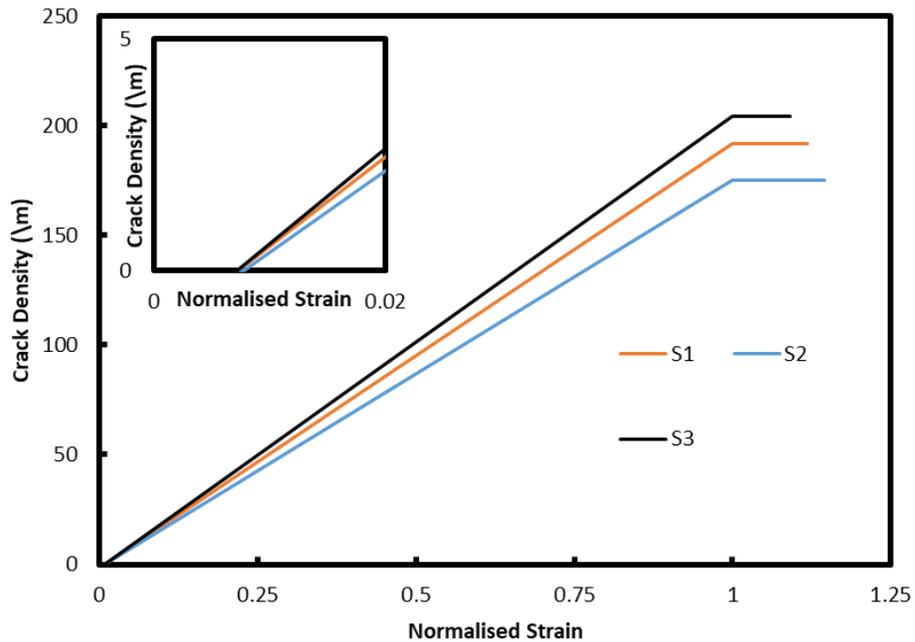

Figure 12 Trilinear estimation of crack density of samples S1, S2, S3

Using the methodology based on TDCNN, CD can be easier obtained from images captured from an existing structure. The trilinear trend of CD vs strain (and more specifically the linear variation over the hardening regime) provides a convenient approach for estimating the degree of damage for prioritization of repair. For the SHCC employed for a particular structure, the CD vs strain relation would be measured in the laboratory first to provide a reference. For a real structural member, the observed CD can be divided by the maximum crack density ($CD_m$) and the ratio would indicate the closeness to failure. From the practical point of view, failure can be taken to occur when CD/CDm is equal to one. As shown in Figure 12, the strain ($\epsilon_{lcr}$) at which maximum CD is reached is below the failure strain when crack localization occurs. As crack localization is very difficult to measure without permanently installed instruments, the use of the above ratio will hence provide a conservative criterion, which is actually advantageous for deciding the urgency of repair.

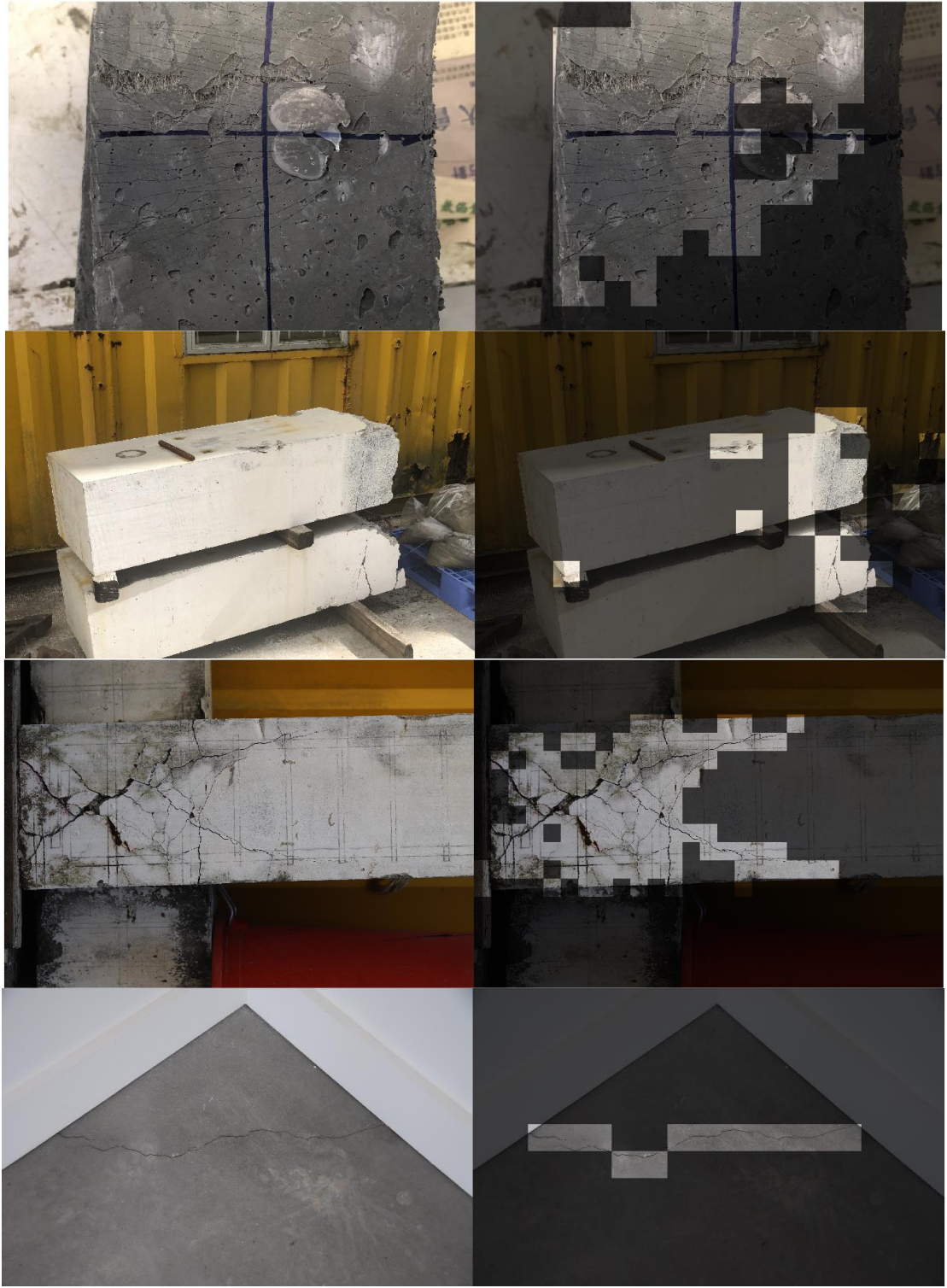

*Figure 13 Examples of crack detection in a natural environment utilizing TDCNN*

5 Additional Studies

5.1 Application of TDCNN for detection and locating cracks in natural environments

Detecting cracks in real structures is challenging probably due to presence of multiple objects in a construction environment. A total of 50 pictures showcasing surface damage in cementitious structures in a natural but challenging environment were collected. Some examples of these images are shown in Figure 13. Following the process described in section 4, the damages present within each of this picture was identified and located utilizing TDCNN. The results are also shown in Figure 13. In these images, positively identified blocks are highlighted. That is the segments of the image where there is no damage is found, are darker than those of with damage (cracks). The performance parameters accuracy and F1-Score for these 50 images are 0.92 and 0.89 respectively. This short study shows even though with limited number of pictures of damages in natural environment, the results observed from this study is a good indication that proposed technique is applicable even in a (challenging) natural environment.

5.2 Explanation behind superior prediction accuracy of TDCCN

Comprehensive understanding of the reasons leading to the superior performance of TDCCN is essential for the scientific progress of this technique. To this end, a short study has been performed to understand the working mechanism of DCCN. Many researchers have proposed techniques to rate and quantify interpretability for a better understanding of the inner mechanisms of (deep) convolutional neural networks [62], [63]. One of the most widely adopted techniques is to visualize the first-layer weights of trained DCNNs [34]. Usually, the features learned in the first layer of DCNNs is very abstract and do not resonate well with the human understanding of such interpretation [62]. Nonetheless this technique can still be used to uncover the quality of training by visually comparing a poor performing and a good performing DCNNs for a specific task [34]. The interpretability measured as closeness between a human interpretation of the concept and machine representation of the concept increases with deeper layers [62]. Therefore, to understand the explainability of the DCCN, the activations of the last convolutional unit of the TDCCN were investigated. The objective of designing TDCCN is to segment image with crack(s) (i.e. P) from the negative images, therefore, in this section, only selected P annotated images (i.e. images which has cracks) were selected. For this study, 100 P images were selected randomly from NNC and SCC datasets. This new dataset is called the explainability (Expla) dataset. Some of the images of Expla dataset are shown in Figure 14. For each image of Expla dataset ($I_E$), the activation map $A_f(I_E)$ of the final convolution layer (f) was computed. Since the resolution of the input image is shrunk (lower than the input image) with each convolutional layer effective comparison of a low-resolution $A_f(I_E)$ to the resolution of the input image $I_E$ is performed after $A_f(I_E)$ is scaled up using bilinear interpolation following the process described in [62]. As shown in Figure 14, the $A_f(I_E)$ for each input image is overlaid on the corresponding image as a heatmap. The highest weight in each heatmap is shown in red and the lowest as transparent. The results in figure 14 show qualitatively that the discriminative power of TDCCN is attributed to correct activation (weightage) of the pixels where crack exists within an image. This behavior is consistent with human understanding /identification of the cracks.

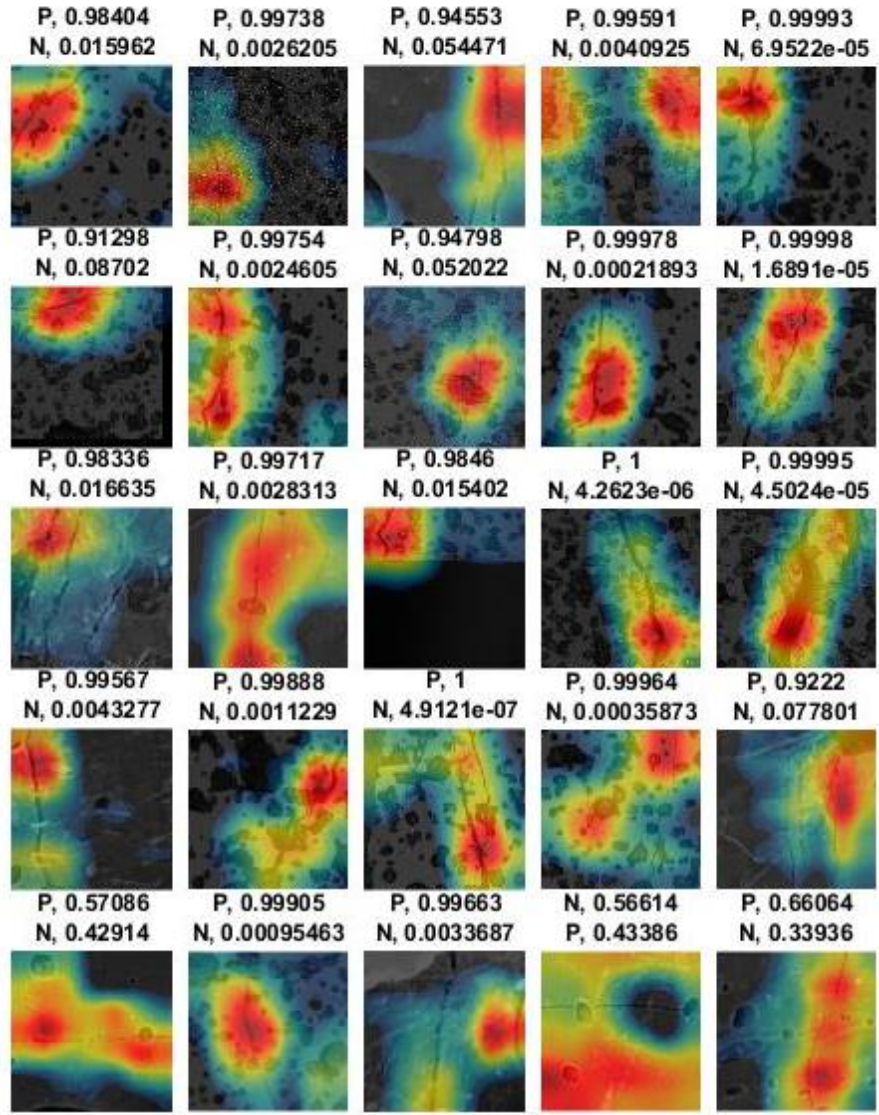

Figure 14 Heatmap showcasing regional influence within the picture for crack classification through TDCCN. Note: heatmap is overlaid over original image for better visulization

6 Conclusion

   Strain Hardening Cementiitous Composites which exhibit large tensile ductility (2-3%) through intrinsically controlled multiple fine cracks (widths ~100 microns) are being utilized to create earthquake resistant component and durable structures. Identification and computation of surface cracks parameters of SHCCs could potentially uncover the details of the operational health and facilitate durability design in a practical environment. Screening of large structural elements for qualitative identification of potentially unsafe components is useful for proper repair (re-strengthening) after a hazard has occurred. In this work, we have presented to systematic approach to explore the potential of deep convolutional networks for rapid identification cracks in SHCC.

- A novel cross domain application i.e. detection of thin multiple cracks of SHCC using pre-trained deep convolutional network is demonstrated.
- Results indicate that tailored deep convolutional network (TDCNN) is invariant against epistemic changes in the image quality, could adapt to a new adverse dataset in the future, and is capable of a human-level inference. This demonstrates the robustness and superiority of TDCNN as compared to other traditional techniques.
- It was observed that even when the images collected is not optimal, by addition of these images in the dataset improves the performance of TDCNN this, will be beneficial for field application where, stability of the camera hardware can not be always guaranteed.
- A image processing technique to extract crack features- crack density (CD) from the results of TDCNN was proposed. The CD computed using TDCNN increases in the hardening regime to a maximum value and remains somewhat constant before the ultimate strength is reached. By comparing the crack density of a member with the maximum value the level of deterioration could be estimated.

Through this research a novel trained ConvNet-TDCNN was trained which can classify thin cracks in SHCC even in presence of artifacts such as sensors (LVDTs) sensor wires, markings, uneven sample edges, background cluttered samples and presence of additional objects found in and around laboratory (such as setups used in loading, residue from glues) which has similar features as cracks. In the future, instead of incorporating an additional image processing technique we would work towards integrating damage severity index within the TDCNN by changing the last layers. This will allow a end-to-end result which will be more convenient for engineers than the current version. Moreover, by collecting additional dataset which showcases artifacts similar to in field conditions the current version of the neural network-TDCNN could be more easily integrated with practice.

Appendix 1: Performance Parameters

$$Accuracy = (TP + TN)/(FP + TP + TN + FN) \tag{A.1}$$

$$Precision = TP/(TP + FP) \tag{A.2}$$

$$Recall = TN/(TN + FN) \tag{A.3}$$

$$F1\ Score = 2 * Precision * Recall/(Precision + Recall) \tag{A.4}$$

here, TP is true positive, FP is false positive, TN is true negative, and FN is false negative

Appendix 2: Intuition behind receiver operating characteristic (ROC) curve
A and B are 2 classes of images each containing N images (objects) (I).

$$A = [I_{A1} \ldots \ldots I_{AN}]\ ,\ B = [I_{B1} \ldots \ldots I_{BN}] \tag{A2.1}$$

A classifier is expected to decompose such images along a feature space to distinguish sets of images B from A. Depending the quality of the feature space, the decomposed images of each class have a distribution $D_A$ and $D_B$ respectively. $D_A$ and $D_B$ are either completely separate, partially separated and completely overlapped as shown in Figure A2.1, respectively.

To quantify the quality of the separability, threshold value (T) was slided within the range of $D_A \cup D_B$ along x axis. That is LA is lower limit and RB is the upper limit for T. For each value of T, true positive rate

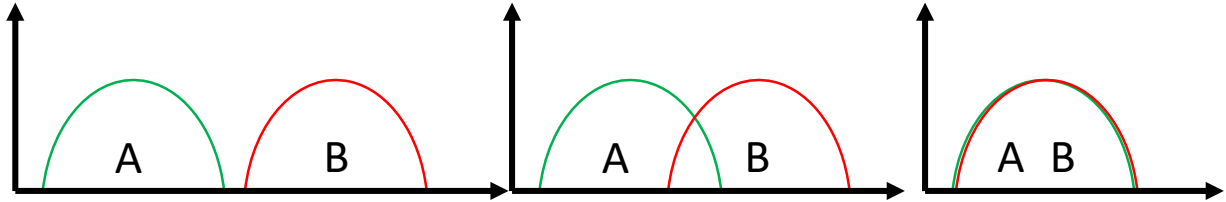

*A2 1 Possible distribution of classes of image along feature space a) well separated b) partially separated c) overlapped*

(TPR) and false positive rate (FPR) according to Eqn. A.2.2 and A.2.3, respectively were computed.

$$TPR = \int_{LA}^{T} D_A \qquad (A.2.2)$$

$$FPR = \int_{LA}^{T} D_B \qquad (A.2.3)$$

By plotting, TPR against FPR values for different values of threshold ROC curve of a classifier was estimated. Figure A2.2 shows the typical ROC curve for the cases introduced in Figure A2.1. A more

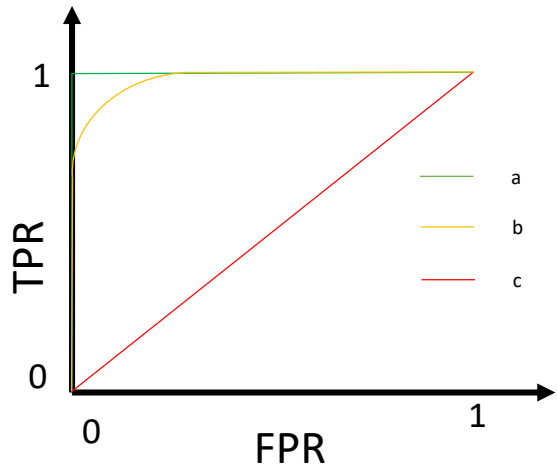

*A2 2 ROC curve for class A for cases a, b,c with reference to figure A2 1*

detailed description on ROC curve can be found in reference [55]. Expectedly, area under ROC curve can be used as indicator for ranking the performance of classifiers.

Appendix 3: Computation of crack pattern and crack number from LCZ

Computation of crack statistics involves the following 4 steps. In the first step, LCZ within an image using DCCN was computed as described in section 4.2 . One example is shown in Figure A3. 1a. Within the LCZ, (enhanced) image segments corresponding to a window which is classified as P was extracted. This selection is limited to positive windows which has at least one neighbors. One such example is marked in green in Figure A3 1a and enlarged version is shown in A3 1b. However, for windows with no

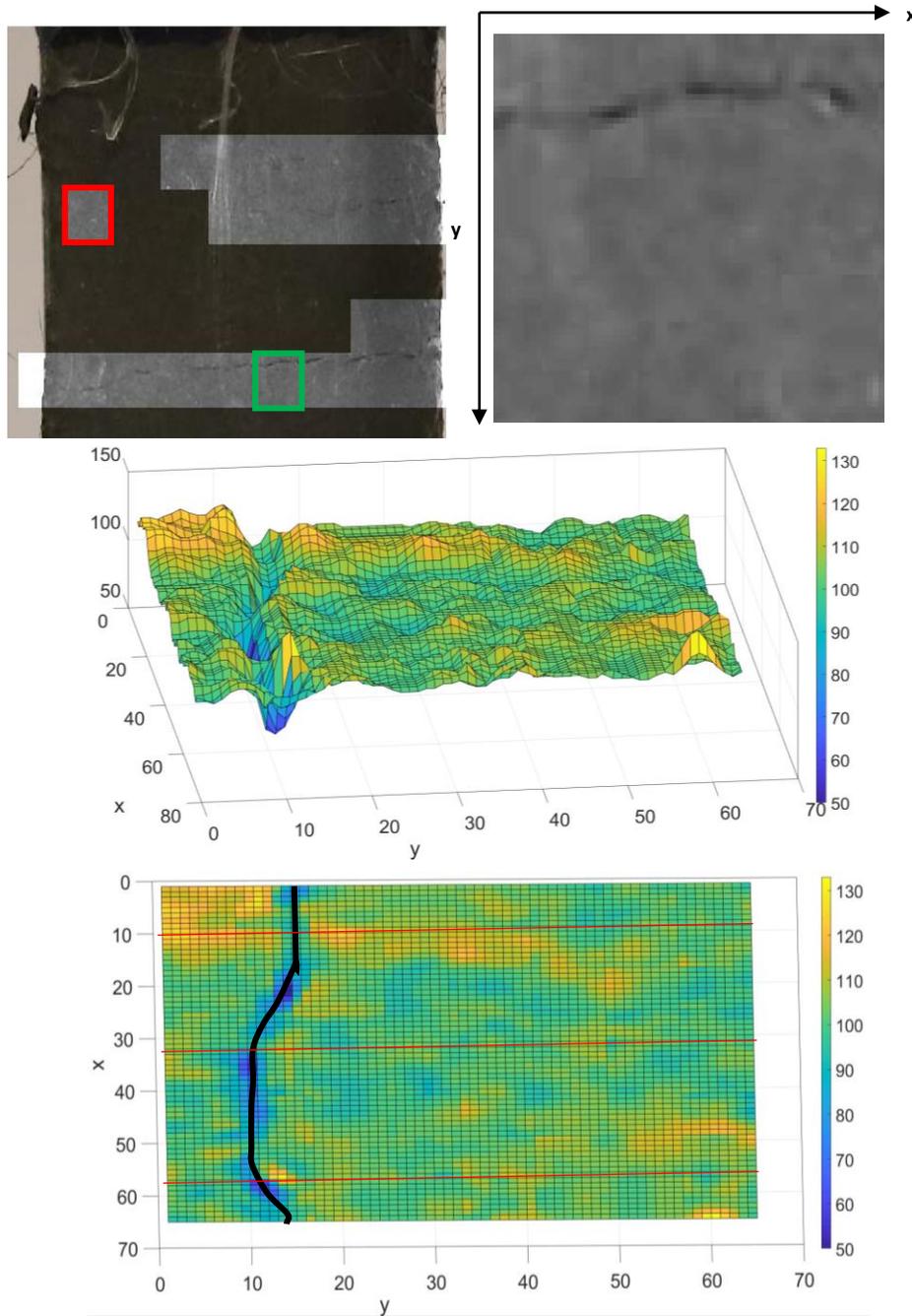

A3 1 Schematic representation for computation of crack statistics a) Typical LCZs within a frame b)Selected window c) Intensity distribution within an image segment d) Crack Length and crack number calculation (Note: crack is annotated with black line and scanning lines are annotated as red) neighbors such as one marked in red in Figure A3 1a, were not analyzed them further with the assumption that a crack is going to be at least larger than one window. On the other hand for A3 1b (and others within green boxes) intensity distribution was calculated (shown in A3 1c). Within the surface plot, followed the center of (a local) through created by dark pixels was followed to annotate crack within the window. This is schematically shown through figure A3 1d. By combining all such annotated cracks, crack pattern was estimated within each image. The length of the annotated line segment is thus defined as crack length. The crack number is calculated by averaging the number of times cracks crosses scanning lines (marked in red in figure A3 1d).